\documentclass[12pt, oneside]{article}

\usepackage[paper=a4paper, total={16cm,23cm}]{geometry}
\usepackage[T1]{fontenc}
\usepackage{setspace} 
\usepackage[hang]{footmisc} 
\usepackage{titling} 
\usepackage{tocloft} 
\usepackage{parskip} 
\usepackage{amsfonts}
\usepackage{amsmath}
\usepackage{amssymb}
\usepackage{mathrsfs, dsfont, amsthm}
\usepackage{physics}
\usepackage{mathtools}
\usepackage{bbm}
\numberwithin{equation}{section} 

\usepackage[dvipsnames]{xcolor}

\usepackage{hyperref} 
\definecolor{Gary}{RGB}{100, 100, 100}
\hypersetup{colorlinks=true, allcolors=Gary, linktoc=all}

\usepackage{cite}

\usepackage{tikz}
\usepackage{tikzlings}
\usetikzlibrary{calc, patterns, decorations, decorations.markings, shapes, positioning, intersections, quotes, angles}
\usepackage[subrefformat=parens]{subcaption}
\usepackage{pgfplots}
\pgfplotsset{compat=newest}
\newcommand{\mathdefault}[1][]{}

\newtheorem*{definition*}{Definition}
\newtheorem*{claim*}{Claim}
\newtheorem*{conjecture*}{Conjecture}

\begin{document}
\begin{titlingpage}
    \vspace*{10em}
    \onehalfspacing
    \begin{center}
    {\bf \Large Crossing symmetry of OPE statistics}
    \end{center}
    \singlespacing
    \vspace*{2em}
    \begin{center}
        \textbf{
        Diandian Wang
        }
    \end{center}
    \vspace*{1em}
    \begin{center}
        \textsl{
        Center for the Fundamental Laws of Nature,\\
        Harvard University, 
        Cambridge, MA 02138, USA\\[\baselineskip]
        }
        \href{mailto:diandianwang@fas.harvard.edu}{\small diandianwang@fas.harvard.edu}
    \end{center}
    \vspace*{3em}
    \begin{abstract}
    We study the crossing symmetry of an ensemble of large-$c$ 2D CFTs defined through 3D gravity. A central observation is that statistical moments of OPE coefficients are not independent; rather, lower and higher moments are strongly correlated. Using Virasoro TQFT, we clarify how these correlations arise and how they guarantee consistency across OPE channels. Our analysis introduces the new notion of pseudo-hyperbolic manifolds, which are a certain class of non-hyperbolic manifolds whose partition functions are nevertheless related to those of hyperbolic ones. These manifolds serve as bridges that help manifest the crossing symmetry of the CFT ensemble.
    \end{abstract}
\end{titlingpage}
\tableofcontents

\section{Introduction}

In a CFT, crossing symmetry requires that a correlation function be independent of the choice of OPE channel, namely that different ways of fusing operators in a correlator must give the same result. This reflects the associativity of the OPE and strongly constrains the spectrum and OPE coefficients. 

Recent years have seen the emergence of a novel relationship between gravity in lower dimensions and ensembles of theories, which has attracted considerable interest. For instance, the presence of Euclidean wormholes in 3D gravity suggests a putative dual boundary description as an ensemble of (approximate) CFTs \cite{CJ20,DUP23,BDFPR25,BdB20,BdBL21,ABdBL21,CCHM22,CEZ23,BdBJNS23,dBLPS23,CEZ24,dBLP24,JRW24,Chandra25,Hartman25b}. (See also \cite{CEMR23,JRSW25,WWW25,HJL25,JRW25} for ensembles of BCFTs.) However, individual draws from the ensemble do not in general have crossing symmetry. It is therefore unclear whether the CFTs in the ensemble collectively possess crossing symmetry, or what that even means. 

To make progress, we recall the definition of the ensemble. Take a product of correlation functions and take the average. This is called a multi-copy observable \cite{CCHM22}. In an ensemble, the average of a product does not generally factorize into the product of the averages. Computing this is complicated in general, as this object depends on the OPE coefficients and the spectrum, both being random variables. Moreover, the correlation function depends on the conformal weights in a non-algebraic manner. A tensor model has been designed to deal with this general situation \cite{BdBJNS23,JRW24}. 

However, our goal is less ambitious as we work at large $c$ where things simplify significantly. For one thing, the wormhole that signals the random matrix nature of the spectrum is order one in $c$ \cite{CJ20}, allowing us to treat the spectrum as fixed rather than random. Computing multi-copy observables then reduces to a more manageable problem. As the correlation functions depend on the OPE coefficients algebraically, conformal blocks can essentially be moved outside of the average. One therefore only needs to know about quantities that are of the schematic form $\overline{C_{ijk}^n}$, where $C_{ijk}$ denotes an OPE coefficient and the overline denotes averaging. These are multivariate statistical moments for the OPE coefficients, or OPE moments for short. 

While defining an ensemble from an intrinsically CFT point of view has been attempted \cite{BdBJNS23,BMS25}, establishing the agreement with 3D gravity is still an effort in progress.\footnote{Certain OPE moments are in some sense universal and obtainable from bootstrap \cite{CMMT19,ABdBL21}, and the corresponding gravity answers are known to agree with the bootstrap results \cite{CCHM22,BdBL21}. In these situations, the agreement can be understood by uplifting the bootstrap procedure to 3D, as explained in \cite{WWW25}, Section~8.4.}  We circumvent this issue by defining the ensemble via 3D gravity. To do this, we follow \cite{CEZ23,CEZ24}. The OPE moment is defined by finding wormholes connecting asymptotic boundaries with the topology of a three-punctured sphere. Each three-punctured sphere represents an OPE coefficient. In the bulk, Virasoro Wilson lines connect the asymptotic punctures with the same labels. In the large-$c$ limit, only certain manifolds with Wilson lines inserted in certain ways contribute. These are hyperbolic manifolds, defined to be the set of manifolds with on-shell configurations. We should emphasize that the Wilson line configuration is part of the definition of hyperbolicity, as Virasoro Wilson lines encode topological information (unlike matter Wilson lines). Geometrically, they become spacetime bananas \cite{AAMV23a,AAMV23b} at large $c$ if the states connected by the Wilson lines are above the black hole threshold, or conical defects if they are below the threshold \cite{DJ84,Krasnov00}. In this work, the ensemble does not include any states below the black hole threshold except the vacuum.

Now that we have a definition of the OPE moments, we can go back to the multi-copy observables and pose the question of crossing symmetry: do we get the same answer for a multi-copy observable regardless of the choice of the OPE channel for each copy? From the perspective of constructing the ensemble, this is a strong requirement. For example, it forces the ensemble to be non-Gaussian \cite{BdBL21}. From the gravity perspective, however, this seems intuitive: diffeomorphism of gravity should ensure that no channel is preferred. The goal of the current work is to establish this explicitly and to clarify several points that will help manifest the crossing symmetry. 

The structure of the rest of the paper is as follows. In Section~\ref{sec:rev}, we review an example of a multi-copy observable in the CFT ensemble and demonstrate its crossing symmetry explicitly. In Section~\ref{sec:pseudo}, we define the notion of pseudo-hyperbolicity, which will then play an important role in Section~\ref{sec:cross} where we give an argument for establishing the crossing symmetry. Finally, we provide some comments in Section~\ref{eq:disc}.

\paragraph{Note added.} The idea of ``two-point joining'' was arrived at independently by the authors of \cite{BCELP26}, who first presented it publicly in several talks before the submission of version~1 of the arXiv preprint of this work.

\section{Review of the CFT ensemble}\label{sec:rev}

Let us begin by reviewing an example studied in \cite{CCHM22}. Consider the product of two four-point functions on the sphere. Write this as
\begin{align}
G_{1234}^*G_{1234}^{\prime }\equiv\left\langle\mathcal{O}_1(0) \mathcal{O}_2(z, \bar{z}) \mathcal{O}_3(1) \mathcal{O}_4(\infty)\right\rangle
^*\left\langle\mathcal{O}_1(0) \mathcal{O}_2(z^{\prime}, \bar{z}^{\prime}) \mathcal{O}_3(1) \mathcal{O}_4(\infty)\right\rangle.
\end{align}
In contrast to \cite{CCHM22}, all operators we consider here are above the black hole threshold and have general spin. Without loss of generality, expand the first four-point function in the $12\to p\to 34$ channel, i.e., the $s$ channel. For the second four-point function, we can either expand it in the same channel or a different channel. Let us now compare the two options. 

If the second copy is also expanded in the $s$ channel, this object then evaluates to
\begin{align}\label{eq:GGs}
\left.\overline{G_{1234}^* G_{1234}^{\prime}}\right|_{s}=\sum_{p, q} \overline{C_{12 p}^* C_{34 p}^* C_{12 q} C_{34 q}}\left|\mathcal{F}_{1234}(P_p ; z) \mathcal{F}_{1234}(P_q ; z^{\prime})\right|^2,
\end{align}
where $\mathcal{F}_{1234}(h_p;z)$ is the (holomorphic) Virasoro conformal block with internal weight $h_p$ and cross ratio $z$, and $|\cdot|^2$ denotes multiplication by the anti-holomorphic counterpart. Following \cite{CCHM22,JRW24}, we take the spectrum to be given by the Cardy density above the black hole threshold plus the vacuum, so the sum actually means
\begin{align}\label{eq:sumdef}
    \sum_p \longrightarrow \int_0^{\infty} \mathrm{d}P_p\mathrm{d}\bar{P}_p\,\left[\rho_0(P_p)\rho_0(\bar{P}_p)+\delta(P_p-\mathbbm{1})\delta(\bar{P}_p-\mathbbm{1})\right].
\end{align}

To find $\overline{C_{12 p}^* C_{34 p}^* C_{12 q} C_{34 q}}$, we look for 3D manifolds that have four spherical boundaries, each having three punctures with the labels corresponding to the subscripts of the OPE coefficients. The punctures on the boundaries are connected by Wilson lines in the bulk.

One contribution to this is given by a disconnected term,
\begin{align}\label{eq:4ptgauss}
\overline{C_{12 p}^* C_{34 p}^* C_{12 q} C_{34 q}} \supset \overline{C_{12 p}^* C_{12 q}}~\overline{C_{34 p}^* C_{34 q}}\supset|C_0(12p)|^2|C_0(34q)|^2\frac{\delta^2({P_p-P_q})}{|\rho_0(p)|^2},
\end{align} 
where $C_0$ is the DOZZ formula \cite{DO94,ZZ95} in the CMMT normalization \cite{CMMT19}, and we have used a simplified notation
\begin{align}
    C_0(P_i,P_j,P_k)\to C_0(ijk),\quad \rho_0(P_i)\to \rho_0(i).
\end{align}
Similar notational simplifications will be employed when Liouville momenta appear in other functions as well. The contribution \eqref{eq:4ptgauss} is also known as the (leading) Gaussian contraction, and it corresponds to the so-called $C_0$ wormhole:
\begin{align}\label{eq:C0}
    \overline{|C_{ijk}|^2}\supset \vcenter{\hbox{\includegraphics[height=1.5cm]{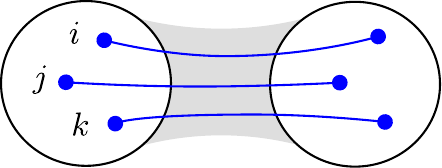}}}=|C_0(ijk)|^2.
\end{align}
Here and henceforth, the drawing of the manifold is usually used to mean its Virasoro TQFT partition function. The punctured spheres in this figure depict asymptotic boundaries. 

Another contribution comes from
\begin{align}\label{eq:samechannelsecond}
\overline{C_{12 p}^* C_{34 p}^* C_{12 q} C_{34 q}} &\supset 
\overline{C_{12 p}^* C_{43 p}}~\overline{C_{12 q} C_{43 q}^*}
\nonumber
\\
&\supset|C_0(12p)|^2|C_0(34q)|^2
\frac{\delta^2({P_1-P_4})}{|\rho_0(1)|^2}
\frac{\delta^2({P_2-P_3})}{|\rho_0(2)|^2},
\end{align} 
which is a different way of pairing up the indices. Note that we have used $C_{ijk}^*=C_{jik}$ in the first line.

If the second copy is expanded in the $t$ channel instead, it becomes
\begin{align}\label{eq:GGt}
\left.\overline{G_{1234}^* G_{1234}^{\prime }}\right|_{t}=\sum_{p, q} \overline{C_{12 p}^* C_{34 p}^* C_{41 q} C_{23 q}}\left|\mathcal{F}_{1234}(P_p ; z) \mathcal{F}_{4123}(P_q ; z^{\prime})\right|^2.
\end{align}
With this index structure for the OPE moment, one contribution is a four-boundary wormhole, the so-called $6j$ manifold:
\begin{align}\label{eq:6jexpression}
    \overline{C_{p21} C_{p43} C_{41 q} C_{23 q}}&\supset\vcenter{\hbox{\includegraphics[height=4cm]{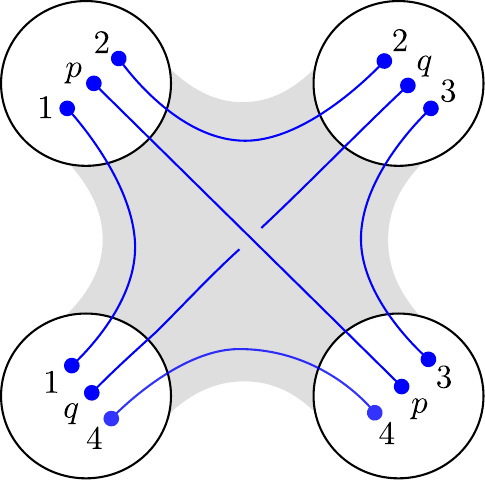}}}
    \nonumber\\
    &=\sqrt{|C_0(12p) C_0(34p) C_0(14q) C_0(23q)|^2}
    \left|\left\{\begin{array}{lll}
    q & 4 & 1 \\
    p & 2 & 3
    \end{array}\right\}\right|^2,
\end{align}
where the $6j$ symbol here has the Racah-Wigner normalization \cite{TV12,TV13,Eberhardt23}.

Crossing symmetry requires that the answer be independent of the OPE channel in which we expand each copy of the observable. To see what this implies for the statistical moments, first write the $t$-channel block in terms of the $s$-channel block,
\begin{align}
    \mathcal{F}_{4123}(P_q;z')
    =\int_0^\infty \mathrm{d} P_{q'}
    \,
    \mathbb{F}_{q{q'}}\!
    \begin{bmatrix}
    3 & 4 \\
    2 & 1
    \end{bmatrix}\mathcal{F}_{1234}(P_{q'};z'),
\end{align}
where $\mathbb{F}$ is the spherical crossing kernel, also known as the fusion matrix.
Comparing \eqref{eq:GGs} and \eqref{eq:GGt}, we see that crossing symmetry requires
\begin{align}\label{eq:condint}
&\sum_{p, q} \overline{C_{12 p}^* C_{34 p}^* C_{12 q} C_{34 q}}\left|\mathcal{F}_{1234}(P_p ; z) \mathcal{F}_{1234}(P_q ; z^{\prime})\right|^2
\nonumber\\
=\,&
\sum_{p, q}
\int_0^\infty \mathrm{d}^2 P_{q'}
    \,
    \left|\mathbb{F}_{q{q'}}\!
    \begin{bmatrix}
    3 & 4 \\
    2 & 1
    \end{bmatrix}
    \right|^2
    \overline{C_{12 p}^* C_{34 p}^* C_{41 q} C_{23 q}}\left|\mathcal{F}_{1234}(P_p ; z)\mathcal{F}_{1234}(P_{q'};z') \right|^2
    \nonumber\\
=\,&
\sum_{p,q'}\int_0^\infty \mathrm{d}^2 P_{q}
    \left|\mathbb{F}_{q'{q}}\!
    \begin{bmatrix}
    3 & 4 \\
    2 & 1
    \end{bmatrix}
    \right|^2
    \overline{C_{12 p}^* C_{34 p}^* C_{41 q'} C_{23 q'}}\left|\mathcal{F}_{1234}(P_p ; z)\mathcal{F}_{1234}(P_{q};z') \right|^2,
\end{align}
where we have relabeled $q\leftrightarrow q'$ in arriving at the last line.

For this to hold, it is sufficient to show that
\begin{align}
\overline{C_{12 p}^* C_{34 p}^* C_{12 q} C_{34 q}}
\stackrel{!}{=}\,
&
\sum_{q'}\frac{1}{|\rho_0(q)|^2}
    \left|\mathbb{F}_{q'{q}}\!
    \begin{bmatrix}
    3 & 4 \\
    2 & 1
    \end{bmatrix}
    \right|^2
    \overline{C_{12 p}^* C_{34 p}^* C_{41 q'} C_{23 q'}},
\end{align}
where $P_q$ is the weight of any state in the spectrum. For simplicity, we restrict ourselves to considering real $P_q$ in this example, and we will discuss the case with $P_q=\mathbbm{1}$ in Section~\ref{ssec:gensphere} where we study the general case.

Applying the definition \eqref{eq:sumdef}, it reads
\begin{align}\label{eq:condition}
\overline{C_{12 p}^* C_{34 p}^* C_{12 q} C_{34 q}}
\stackrel{!}{=}\,
&
\int\mathrm{d}^2P_{q'}
    \frac{|\rho_0(q')|^2}{|\rho_0(q)|^2}\left|\mathbb{F}_{q'{q}}\!
    \begin{bmatrix}
    3 & 4 \\
    2 & 1
    \end{bmatrix}
    \right|^2
    \overline{C_{12 p}^* C_{34 p}^* C_{41 q'} C_{23 q'}}
\nonumber\\
+\,&
    \left|C_0(12q)
    \right|^2
    \overline{C_{12 p}^* C_{34 p}^*}
    \frac{\delta^2(P_1-P_4)}{|\rho_0(1)|^2}
    \frac{\delta^2(P_2-P_3)}{|\rho_0(2)|^2},
\end{align}
where we have used
\begin{align}
\mathbb{F}_{\mathbbm{1}q}
\begin{bmatrix}
    2 & 1 \\
    2 & 1
    \end{bmatrix}
=C_0(12q)\rho_0(q)
\end{align}
and the normalization of the two-point function
\begin{align}
    C_{ij\mathbbm{1}}=\frac{\delta^2(P_i-P_j)}{|\rho_0(i)|^2}.
\end{align}

Substituting the contributions \eqref{eq:4ptgauss} and \eqref{eq:samechannelsecond} into the LHS, \eqref{eq:6jexpression} and \eqref{eq:C0} into the RHS of \eqref{eq:condition}, and evaluating the integral, one can check explicitly that the equation holds as far as these contributions are considered. 

More generally, we need to consider multi-copy observables and can independently choose a channel for each copy. We then look for all hyperbolic manifolds with three-punctured sphere boundaries with Wilson line labels matching the index structure of the OPE coefficients. Crossing symmetry of the ensemble is the statement that we will get the same answer regardless of the choice of the channels. Before examining this further, it is useful to introduce the concept of pseudo-hyperbolicity.

\section{Pseudo-hyperbolicity}\label{sec:pseudo}

On hyperbolic manifolds, Virasoro TQFT computes finite gravitational path integrals \cite{CEZ23}. When working at large $c$, where we use saddle point approximation, only hyperbolic manifolds contribute. Nevertheless, we will identify a class of manifolds that are non-hyperbolic but closely related to hyperbolic ones.\footnote{\label{ft:hyper}We work in the momentum basis, so the definition of hyperbolicity here is about whether the manifold admits a saddle with fixed momentum boundary conditions at the Wilson lines. Since the definition is topological, we can start with either real $P=\bar{P}$ near $P=0$, in which case we can use the boundary conditions defined by Hartman \cite{Hartman25b}, or imaginary $P=\bar{P}$ near $P=0$, in which case we impose conical boundary conditions and can use softwares such  as $\mathsf{Orb}$ to test the existence of the saddle \cite{heard2005computation}.} This will play a key role in the rest of the paper when discussing crossing symmetry. 

Let us define a manifold with Wilson line insertions to be \emph{pseudo-hyperbolic} if it contains at least one one-punctured or two-punctured sphere cross-section such that splitting along them turns it into a hyperbolic manifold, possibly disconnected. \emph{Splitting} is defined as follows. 
For a one-point cross-section, simply remove the Wilson line going through it, cut along the cross-section, and then glue an empty 3-ball to each side of the cut, i.e., the reverse connected sum. For a two-punctured cross-section, cut along the cross-section and then glue the following object to each side of the cut:
\begin{align}\label{eq:ball2pt}
    \vcenter{\hbox{\includegraphics[height=2.5cm]{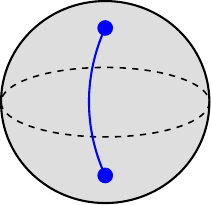}}}~,
\end{align}
which is a 3-ball with a Wilson line extending between two points on the surface. (Gluing two of them to each other makes the Unknot in $S^3$.) 

To see an example, consider the so-called pillow diagram \cite{JRW24},
\begin{align}\label{eq:pillowpic}
    \vcenter{\hbox{\includegraphics[height=4cm]{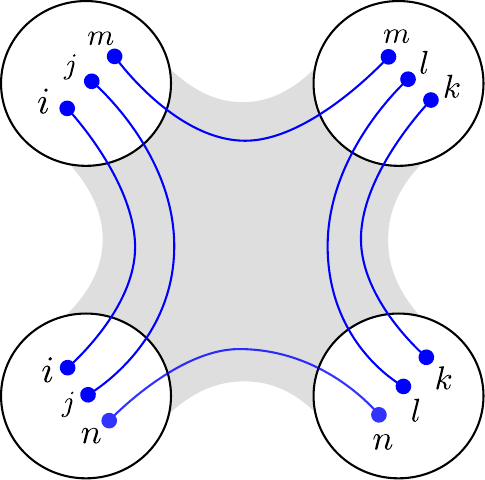}}}~,
\end{align}
which has an associated partition function of
\begin{align}\label{eq:pillow}
   \frac{\delta^2(P_m-P_n)}{\left|\rho_0(m) \right|^2}\left|C_0(ijm) C_0(klm)\right|^2.
\end{align}
This is not hyperbolic, and the partition function is distributional. However, splitting in the middle (along the two-punctured sphere) turns it into two hyperbolic manifolds:
\begin{align}\label{eq:splitpillow}
    \vcenter{\hbox{\includegraphics[height=4cm]{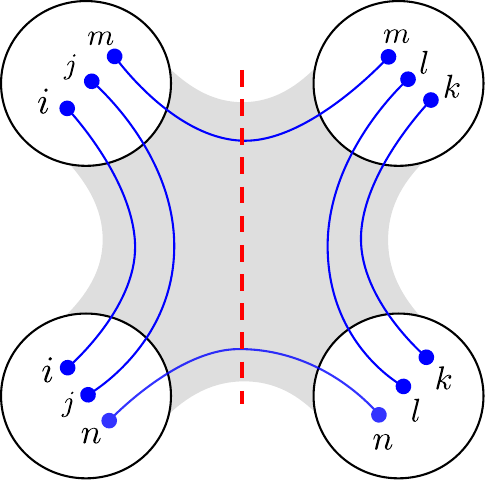}}}
    \quad\longrightarrow\quad
    \vcenter{\hbox{\includegraphics[height=4cm]{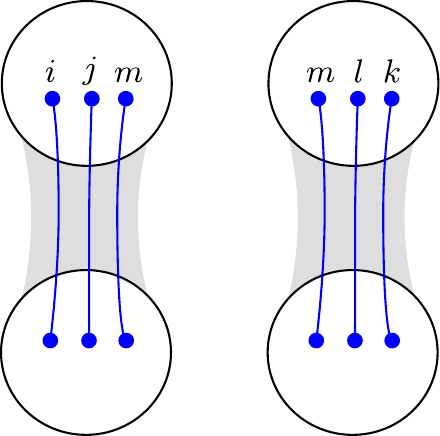}}}~.
\end{align} 
The two manifolds on the right are $C_0$ wormholes, which are hyperbolic. They have partition functions $|C_0(ijm)|^2$ and $|C_0(mlk)|^2$ respectively.

By direct comparison of the partition functions, we observe that the resulting manifold, which is a union of two hyperbolic ones, has the same partition function as the pillow manifold upon removal of a factor
\begin{align}\label{eq:removal}
\frac{\delta^2(P_m-P_n)}{|\rho_0(m)|^2}.
\end{align} 

Now take the pillow diagram \eqref{eq:pillowpic} and set $P_n=\mathbbm{1}$, it then has a one-point cross-section. Using \cite{Eberhardt23}
\begin{align}
\lim_{P_k\to\mathbbm{1}}C_0(ijk)=\frac{\delta(P_i-P_j)}{\rho_0(i)}
\end{align}
to close off the boundaries where the Wilson line labeled by $n$ ends on, this manifold reduces to
\begin{align}
   \vcenter{\hbox{\includegraphics[height=1.5cm]{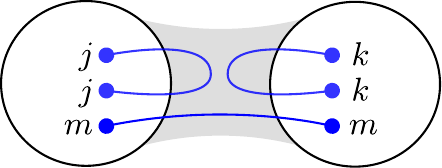}}}=\frac{\delta^2(P_m-{\mathbbm{1}})}{\left|\rho_0(m) \right|^2}.
\end{align}
Splitting along the one-point cross-section then turns it into
\begin{align}\label{eq:onepointsplit}
    \vcenter{\hbox{\includegraphics[height=1.5cm]{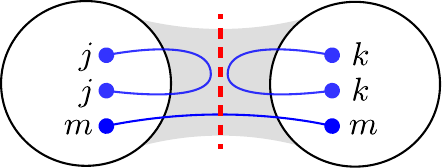}}}
    \quad\longrightarrow\quad
    \vcenter{\hbox{\includegraphics[height=1.5cm]{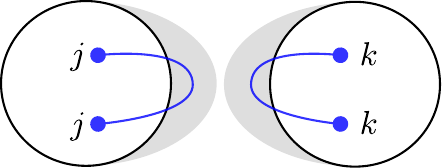}}}~.
\end{align}
As splitting turns this manifold into the union of two hyperbolic manifolds, we say that the manifold (before splitting) is pseudo-hyperbolic. Each manifold on the RHS of \eqref{eq:onepointsplit} has partition function normalized to unity by the normalization of the CFT two-point function. The factor that is removed in this case is therefore
\begin{align}\label{eq:removalonepoint}
    \frac{\delta^2(P_m-\mathbbm{1})}{|\rho_0(m)|^2},
\end{align}
which is a special case of the factor removed in the two-point split \eqref{eq:removal}. The RHS of \eqref{eq:onepointsplit} can also be obtained from the RHS of \eqref{eq:splitpillow} by setting $P_m=\mathbbm{1}$.

Let us examine a second example. Consider the diagram
\begin{align}
    \vcenter{\hbox{\includegraphics[height=1.5cm]{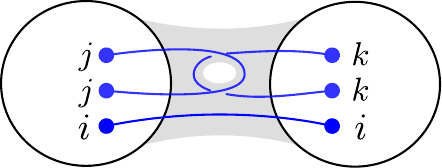}}}~,
\end{align}
which has a genus hole in the middle, preventing the $j$ and $k$ Wilson lines from being deformed to lie entirely on the boundaries. More precisely, if we shrink the two boundary spheres to points to obtain the embedding manifold, the manifold has topology $S^2\times S^1$. The partition function for this manifold is given by \cite{JRW24}
\begin{align}
   \frac{\delta^2(P_j-P_k)}{\left|\rho_0(j) \right|^2}|C_0(i j j)|^2.
\end{align}
This is again pseudo-hyperbolic, as splitting along the two-punctured spherical cross-section leads to a single hyperbolic manifold, which we recognize as the $C_0$ wormhole:
\begin{align}
    \vcenter{\hbox{\includegraphics[height=1.5cm]{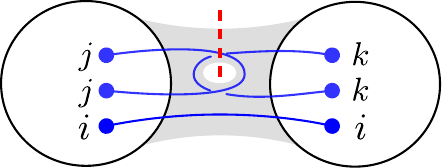}}}
    \quad\longrightarrow
    \quad\vcenter{\hbox{\includegraphics[height=1.5cm]{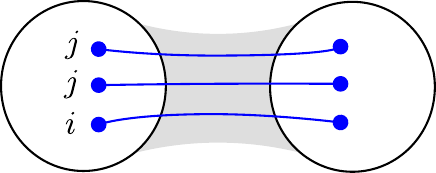}}}~.
\end{align} 
As in the previous example, the partition function of the resulting manifold equals the original one divided by the divergent factor
\begin{align}
    \frac{\delta^2(P_j-P_k)}{|\rho_0(j)|^2}.
\end{align}

More generally, a pseudo-hyperbolic manifold can have an arbitrary number of two-punctured sphere cross-sections. Each such cross-section is associated with a delta function that sets the weights of the two Wilson lines equal \cite{Witten89}. In terms of the partition function, splitting along each such cross-section removes this delta function along with the associated factor of Cardy density. The resulting manifold is free of divergences. By definition, if cutting along such cross-sections does not result in a hyperbolic manifold (possibly disconnected), the original manifold is not pseudo-hyperbolic, i.e., truly off-shell.

Before moving on, let us provide another example. Consider the genus-two Maldacena-Maoz (MM) wormhole \cite{MM04}
\begin{align}
     \vcenter{\hbox{\includegraphics[height=3cm]{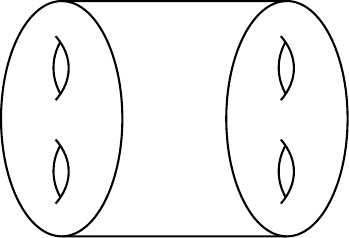}}}~.
\end{align}
A MM wormhole has topology $\Sigma_{g,n} \times I$, where $\Sigma_{g,n}$ is a Riemann surface with genus $g$ and $n$ punctures, and $I$ is an interval.  As they connect two boundaries, they contribute to ensemble-averaged two-copy observables.

Let us first look at the case of paired channels, i.e., if we expand both observables in the same channel. The embedding manifold $M_E$ relevant for computing the correct OPE moment, as defined in \cite{Hartman25b}, can be formed by gluing the MM wormhole to the following two handlebodies with a Wilson graph corresponding to the chosen channel decompositions:
\begin{align}\label{eq:MMg2}
    M_E = \vcenter{\hbox{\includegraphics[height=3cm]{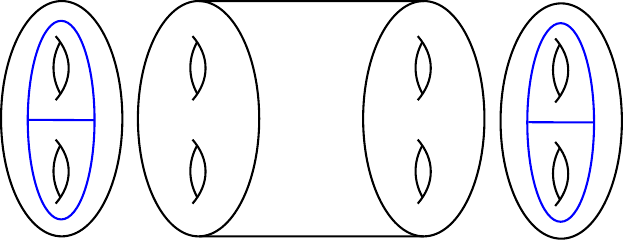}}}~.
\end{align}
The manifold contributing to the OPE moment is obtained by removing the neighborhood of each of the Wilson line junctions, leaving four three-punctured sphere boundaries. For simplicity of the drawing, we will work with the embedding space in this example.

Lengths have no meaning in a topological theory, so imagine shrinking the interval to zero length, i.e., removing the middle piece of the diagram \eqref{eq:MMg2}. The manifold $M_E$ is then just two genus-2 handlebodies, each dressed with the Wilson graph as shown, glued along the genus-2 boundaries as shown. 

A key observation here is that $M_E$ is only pseudo-hyperbolic. This is to be expected, given that each genus-2 handlebody prepares a momentum eigenstate, and the momentum eigenstates form a basis and are therefore orthogonal, with a delta function normalization setting $\vec P=\vec P'$ in the distributional sense.  

To obtain the actual contributing manifolds, we split along all two-punctured spheres (shown in gray):
\begin{align}
 M_E/\mathbb{Z}_2=\vcenter{\hbox{\includegraphics[height=3cm]{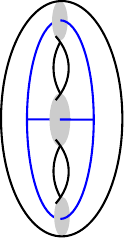}}}~.
\end{align}
For best visualization, we are only drawing half of $M_E$. It is clear that the topology of each cross-section is a two-punctured sphere, as given by doubling the one-punctured disk in $M_E/\mathbb{Z}_2$. Splitting along these three cross-sections turns the manifold $M_E$ into two pieces, each a $C_0$ wormhole (after removing the neighborhoods of the Wilson line junctions). 

This observation is actually general for all MM wormholes. For an arbitrary Riemann surface $\Sigma_{g,n}$, if we expand both copies in the same channel, the splitting procedure above works similarly so that the resulting manifold is always a union of $C_0$ wormholes. This feature underlies the fact that the MM wormhole computes the Liouville partition function on $\Sigma_{g,n}$ because the wormhole direction essentially collapses, and the $C_0$ wormholes play the role of pairs of pants that form the Liouville correlator.

\section{Crossing symmetry of the ensemble}\label{sec:cross}

We are now ready to argue for the crossing symmetry of the ensemble in general. We will start by revisiting the example and providing an explanation from the perspective of 3D gravity. We then move on to establish general crossing symmetry. According to Moore and Seiberg \cite{MS88,MS89}, we only need to establish crossing symmetry of four-point functions on the sphere (spherical crossing) and one-point functions on the torus (modular crossing). General crossing symmetry then follows from the Moore-Seiberg theorem. 

\subsection{Example revisited}

Let us begin by revisiting the example in Section~\ref{sec:rev} and providing an explanation from a perspective that easily generalizes. 

First, we examine both sides of the crossing condition \eqref{eq:condition} from the perspective of Virasoro TQFT. It is useful to rewrite the RHS of \eqref{eq:condition} in terms of $6j$ symbols as
\begin{align}\label{eq:eg-revisit}
&\int\mathrm{d}^2P_{q'}
    \frac{|\rho_0(q')|^2}{|\rho_0(q)|^2}\left|\mathbb{F}_{q'{q}}\!
    \begin{bmatrix}
    3 & 4 \\
    2 & 1
    \end{bmatrix}
    \right|^2
    \overline{C_{12 p}^* C_{34 p}^* C_{41 q'} C_{23 q'}}
\nonumber\\
+\,&
    \left|C_0(12q)
    \right|^2
    \overline{C_{12 p}^* C_{34 p}^*}
    \frac{\delta^2(P_1-P_4)}{|\rho_0(1)|^2}
    \frac{\delta^2(P_2-P_3)}{|\rho_0(2)|^2}
\nonumber\\
\supset\,&{\sqrt{|C_0(12p) C_0(34p) C_0(12q)C_0(34q)|^2}}
    \int \mathrm{d}^2P_{q'}
    |\rho_0({q'})|^2    
    \left|\left\{\begin{array}{lll}
    1 & 2 & q \\
    3 & 4 & q'
    \end{array}\right\}\right|^2\left|\left\{\begin{array}{lll}
    q' & 4 & 1 \\
    p & 2 & 3
    \end{array}\right\}\right|^2
    \nonumber\\
+\,&{\sqrt{|C_0(12p) C_0(34p) C_0(12q)C_0(34q)|^2}}
    \left|\left\{\begin{array}{lll}
    1 & 2 & q \\
    3 & 4 & \mathbbm{1}
    \end{array}\right\}\right|^2\left|\left\{\begin{array}{lll}
    \mathbbm{1} & 4 & 1 \\
    p & 2 & 3
    \end{array}\right\}\right|^2.
\end{align}

In Virasoro TQFT, this has a nice interpretation. In terms of Virasoro TQFT partition functions, multiplication of the two $6j$ symbols above means gluing along the appropriate boundaries:\footnote{In the language of \cite{JRW24}, we need to use the propagator (which is the inverse of $C_0$) whenever we glue two such boundaries. One can also rescale the propagator to 1, at the same time rescaling the normalization of each of the boundaries, as in e.g.~\cite{Hartman25b}. The diagrams here are independent of the normalization, but for concreteness we will always work with the normalization of \cite{CCHM22,CEZ23,JRW24} where all boundaries shown are asymptotic boundaries.}
\begin{align}\label{eq:6j6j}
    \vcenter{\hbox{\includegraphics[height=3.8cm]{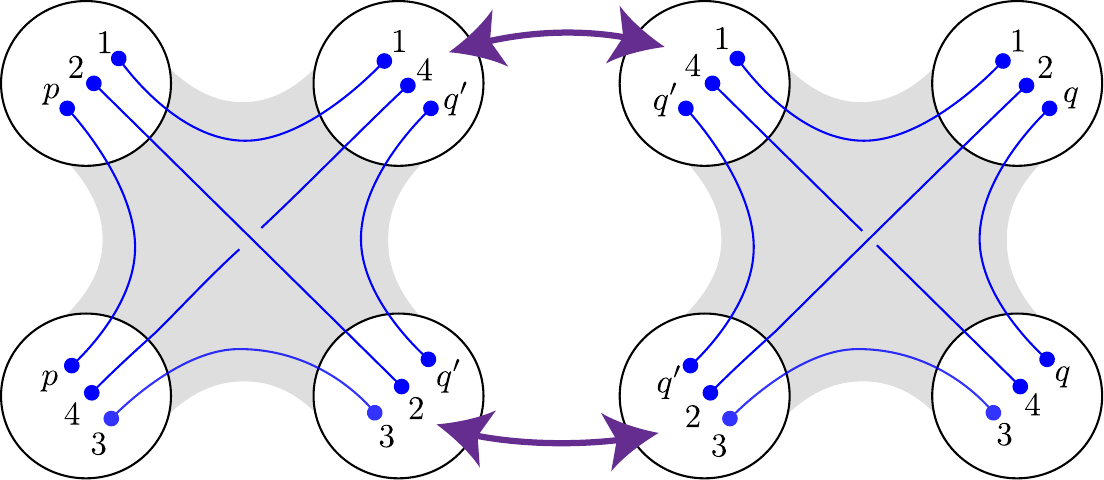}}}\longrightarrow
    \vcenter{\hbox{\includegraphics[height=3.8cm]{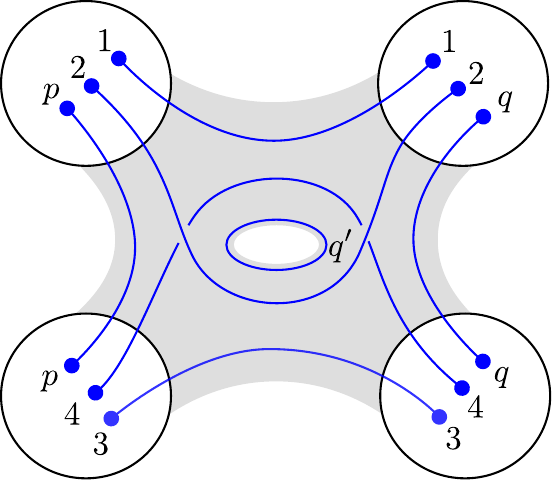}}}~.
\end{align}
This has created a closed loop. For the first term on the RHS of \eqref{eq:eg-revisit}, integrating over the weight of the Wilson loop has a geometric interpretation \cite{BS10}: using the fact that the Cardy density is a special component of the modular S matrix, i.e.,\footnote{Unlike \cite{BCELP26}, which uses the Maloney-Witten-Keller sum \cite{MW07,KM14}, this spectrum is not modular invariant.}
\begin{align}
    \int_0^\infty\mathrm{d}^2P_{q'}\,|\rho_0(P_{q'})|^2\,(\cdot)=
    \int_0^\infty\mathrm{d}^2P_{q'}\,|\mathbb{S}_{\mathbbm{1}q'}[\mathbbm{1}]|^2\,(\cdot),
\end{align}
we can implement this integral via a surgical process on the 3D manifold, known as the \emph{toroidal surgery}.\footnote{It is also naturally referred to as the disk surgery, because this is the disk amplitude in the random matrix description of the spectrum of 3D gravity \cite{CJ20,CEMR23,JRW24,JRW25,JRSW25}. As in JT gravity \cite{SSS19}, only the disk amplitude is on-shell, so we do not need to consider other amplitudes of the matrix model in our situation, which would lead to off-shell manifolds.} To perform the surgery, remove a solid torus neighborhood of the Wilson loop and glue back another solid torus without Wilson lines, with the two cycles swapped. This is the geometric meaning of the S transform. The fact that the solid torus we glue back has no Wilson line is due to the first index of the S matrix being the identity operator. Clearly, this process changes the topology of the manifold, but it can either make the topology more complicated or less complicated. In this particular case, the surgery removes the genus hole, making the topology simpler. Once the genus hole is gone, we can freely deform the Wilson lines, so the resulting manifold is the pillow manifold:
\begin{align}\label{eq:pillowrevisit}
    \vcenter{\hbox{\includegraphics[height=4cm]{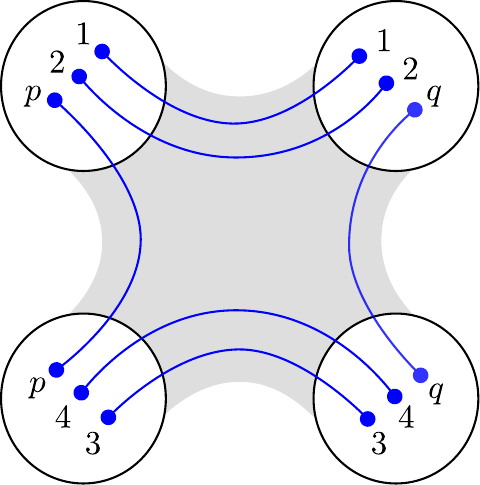}}}~.
\end{align}
In terms of equations, what is implemented by the toroidal surgery in this example is just the orthogonality relation of the Racah-Wigner $6j$ symbols:
\begin{align}
\int_0^{\infty} \mathrm{d} P_r\, \rho_0(r)\left\{\begin{array}{lll}
1 & 2 & q \\
3 & 4 & r
\end{array}\right\}\left\{\begin{array}{lll}
r & 4 & 1 \\
p & 2 & 3
\end{array}\right\}=\frac{\delta(P_p-P_q)}{\rho_0(p)} .
\end{align}
The first term on the RHS of \eqref{eq:eg-revisit} therefore evaluates to
\begin{align}
|C_0(12p) C_0(34p)|^2\frac{\delta^2(P_p-P_q)}{|\rho_0(p)|^2},
\end{align}
which agrees with the corresponding contribution \eqref{eq:4ptgauss} to the LHS of \eqref{eq:condition}. The agreement can be seen from the diagram \eqref{eq:pillowrevisit} after applying the splitting property \eqref{eq:splitpillow}.

For the second term in \eqref{eq:eg-revisit}, the same logic still applies, except that $q'$ in \eqref{eq:6j6j} is now the identity line. Removing it from the diagrams, the resulting manifold no longer contains a closed Wilson loop, and the toroidal surgery is not applicable. However, it now has two two-point cross-sections, so we get to perform splitting twice, resulting in a product of two $C_0$ manifolds, which agrees precisely with the other contribution to the LHS of \eqref{eq:condition}, namely \eqref{eq:samechannelsecond}.

With this diagrammatic way of interpreting the equations, crossing symmetry is now manifest.   Importantly, even though the pillow manifold is a four-boundary wormhole, it is not hyperbolic and is therefore not included as a contribution to the connected statistical moment of $\overline{C_{p21} C_{p43} C_{12 q} C_{34 q}}$. The fact that its partition function is related to two two-boundary wormholes plays a key role in manifesting the crossing symmetry.


The topologies involved might be a little hard to visualize. However, if we consider a $\mathbb{Z}_2$ quotient of the setup, it is possible to describe every step of the process in a visually accurate manner, in the sense that all diagrams can be embedded in $\mathbb{R}^3$ and therefore visualizable by us. For details, see \cite{JRW25} (the $\mathbb{Z}_2$ quotient of the toroidal surgery is called the annular surgery there).

\subsection{General spherical crossing}\label{ssec:gensphere}

In general, spherical crossing is the statement that 
\begin{align}\label{eq:conditiongen}
    \overline{\dots C_{12p} C_{34p}\dots}
    &=
    \sum_{q}
    \frac{1}{|\rho_0(p)|^2}
    \left|\mathbb{F}_{qp}\!
    \begin{bmatrix}
    3 & 4 \\
    2 & 1
    \end{bmatrix}\right|^2
    \overline{\dots C_{41 q} C_{23q}\dots},
\end{align}
where dots include an arbitrary number of other OPE coefficients that appear with the same index structures on both sides, and $P_p$ can either be above the black hole threshold or the identity. In writing this, we assume that all hyperbolic contributions to the OPE moments are included, including both connected and disconnected contributions. 

For every contribution to the statistical moment on the RHS, we can understand the action of the crossing kernel topologically:
\begin{align}\label{eq:glue6j}
    \vcenter{\hbox{\includegraphics[height=4cm]{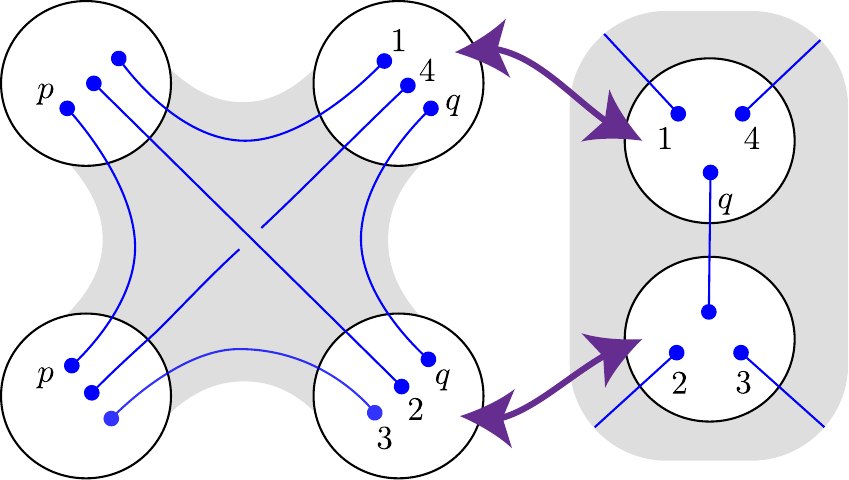}}}~,
\end{align} 
where each purple arrow glues two spheres together. What we have drawn is the neighborhood of the part of the graph that has junctions $41q$ and $23q$ glued to a $6j$ manifold. This procedure generates a genus hole with a Wilson loop wrapping around it (with label $q$ in the picture). If $P_q$ is real and integrated over with the Cardy measure, the toroidal surgery removes this genus hole along with the Wilson loop, as in the example earlier, resulting in
\begin{align}\label{eq:glue6jresult}
    \vcenter{\hbox{\includegraphics[height=3cm]{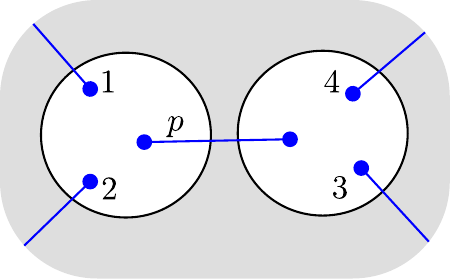}}}~,
\end{align}
which is now a contribution to the LHS of \eqref{eq:conditiongen}. If $P_q=\mathbbm{1}$, we do not do the toroidal surgery and simply set both of the Wilson lines labeled by $q$ in \eqref{eq:glue6j} to the identity: the resulting manifold is pseudo-hyperbolic (as it contains two-point cuts at the gluing spheres) and now contributes to the LHS of \eqref{eq:conditiongen} owing to its index structures being ${12p}$ and ${34p}$. The actual contributions from hyperbolic manifolds are obtained by splitting.

This almost concludes the argument, except that not all contributions to the OPE moment $\overline{\dots C_{41 q} C_{23q}\dots}$ locally look like the picture we drew, the reason being that $C_{41 q}$ and $C_{23q}$ may not belong to the same connected component of the contributing manifold. 

To complete the argument, we simply need to perform the following trick. Whenever we have a contribution to a multi-copy observable that is disconnected, such as
\begin{align}
    \sum_{m,n}\overline{\dots C_{m..} C_{m..}\dots}\,\overline{\dots C_{n..}C_{n..}\dots}
    \,
    \frac{\delta^2(P_m-P_n)}{|\rho_0(m)|^2}\times (\text{conformal blocks}),
\end{align}
we remove a segment of each of the Wilson lines along with a spherical neighborhood containing it and glue a piece that has the topology of a two-punctured sphere times an interval. We call this \emph{two-point joining}, which is the opposite of two-point splitting:
\begin{align}
        \vcenter{\hbox{\includegraphics[height=4cm]{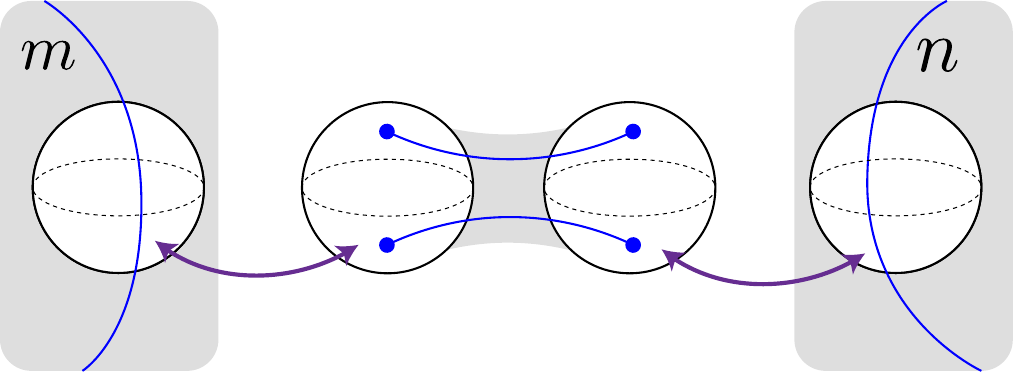}}}~.
\end{align}
The resulting manifold now has a two-punctured sphere cut, so it is pseudo-hyperbolic. If the statistical moment is a product of more than two lower moments, we perform the joining procedure for all such pairings. Notice that the delta functions are nicely absorbed into the partition function of the joined manifold. 

When $P_q=\mathbbm{1}$ on the RHS of \eqref{eq:conditiongen}, we also need to do the opposite of the one-point splitting, which we call \emph{one-point joining}:
\begin{align}
        \vcenter{\hbox{\includegraphics[height=4cm]{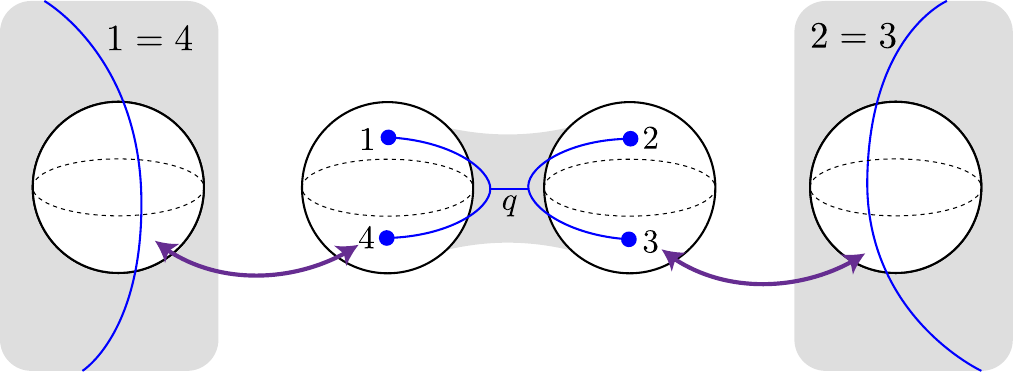}}}~.
\end{align}
Here, to obtain a pseudo-hyperbolic manifold for $\overline{\dots C_{41 q} C_{23q}\dots}$, we start with a contributing manifold (either hyperbolic or pseudo-hyperbolic) for $\overline{\dots \delta_{41} \delta_{23}\dots}$, which is a lower point OPE moment, remove spherical neighborhoods as for the two-point joining, and glue in the object shown. (Also remove spherical neighborhoods of the two vertices.) The Wilson line with label $1=4$ is broken into two lines, now with labels $1$ and $4$ (similarly for $2=3$). We also do the same when other indices in the OPE moments on either side of the crossing equation are the identity. Together with two-point joining, this turns each contribution to the OPE moment into a single manifold, either pseudo-hyperbolic or hyperbolic.

The procedure in \eqref{eq:glue6j} then applies. Performing surgery on this pseudo-hyperbolic manifold turns it into another manifold with different topology, which could either be hyperbolic or pseudo-hyperbolic. If it is pseudo-hyperbolic, we should then split to obtain the actual disconnected contributions to the LHS of \eqref{eq:conditiongen}. This concludes the argument. Notice that this argument applies regardless of whether $p$ is a black hole state or the identity: it at most changes the manifold from hyperbolic to pseudo-hyperbolic. 

To summarize, we have seen that the joining procedure above turns all disconnected contributions to the OPE statistical moments
\begin{align}
    \overline{\dots C_{12p} C_{34p}\dots}
    \quad 
    \text{and}
    \quad
    \overline{\dots C_{41 q} C_{23q}\dots}
\end{align}
into pseudo-hyperbolic manifolds. Therefore, instead of collecting all hyperbolic manifolds, both connected and disconnected, it is equivalent to include only \emph{connected} hyperbolic and pseudo-hyperbolic manifolds when defining the OPE moments.\footnote{The perspective taken in \cite{BCELP26}, where the central object is the partition function rather than the moments, is the same as the latter, because the pseudo-hyperbolic manifolds here do admit saddles after gluing the trumpets that connect them to asymptotic boundaries. See also Footnote~\ref{ft:hyper} to recall the subtle difference.} Establishing \eqref{eq:conditiongen} is then done manifold by manifold. The procedure depicted in \eqref{eq:glue6j} establishes a one-to-one map between the sets of contributing manifolds. (More precisely, the map is bijective, as it is invertible.\footnote{We thank Tom Hartman for making this point.}) The fact that the $6j$ manifold has its partition function given by the fusion matrix then precisely accounts for the extra factor on the RHS of \eqref{eq:conditiongen}.

\subsection{General modular crossing}

We have just seen that spherical crossing symmetry holds for the ensemble. There is a parallel story for the modular crossing kernel. Where the $6j$ manifold was used to implement fusion, we use the following manifold to implement modular crossing:
\begin{align}
    \vcenter{\hbox{\includegraphics[height=1.5cm]{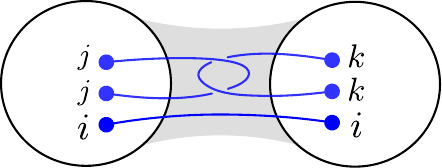}}}
    ~.
\end{align} 
The Virasoro partition function on this manifold is precisely the modular crossing kernel \cite{JRW24}. The method for deriving this is reviewed in \cite{JRW25}.

The statement of modular crossing, which is the analog of \eqref{eq:conditiongen}, is given by
\begin{align}
    \overline{\dots C_{1pp}\dots}
    =
    \sum_{q}\frac{1}{|\rho_0(p)|^2}
    \left|\mathbb{S}_{qp}[1]\right|^2
    \overline{\dots C_{1qq}\dots},
\end{align}
where $\mathbb{S}$ is the modular crossing kernel, and the ``$1$'' labels an arbitrary state above the black hole threshold. (In this case, as $C_{1\mathbbm{1}\mathbbm{1}}=0$, we can replace the sum with an integral over the real $P_q$.)

The situation is now slightly different compared to spherical crossing, as the LHS and RHS have exactly the same index structure. This equality is therefore a statement about just one set of manifolds rather than two. Nevertheless, it is a non-trivial equation. The argument then pairs up the contributing manifolds in a way to manifest the modular symmetry of the sum.

\section{Discussion}\label{eq:disc}

The central concept we introduced in this work is the notion of pseudo-hyperbolicity. In the original work of Virasoro TQFT \cite{CEZ23}, pseudo-hyperbolic manifolds are excluded from consideration, as cross-sections with non-negative Euler characteristic are not allowed (three-punctured spheres are fine, but two-punctured spheres are not). In an ordinary TQFT, a two-punctured sphere cut gives a delta function, and this feature turns out to persist in Virasoro TQFT, with the Kronecker delta replaced by the Dirac delta \cite{JRW24}. It is then tempting to expect that the notion of the connected sum as explained in \cite{Witten89} also generalizes. The na\"ive generalization turns out to be incorrect, as one would need to \emph{divide} by the delta function, which works only for the Kronecker delta. We propose that the splitting property we observed in this work is the correct generalization, though it is unclear what TQFT axiom it follows from, if any. Regardless, this is a useful fact for understanding gravity, and it would be interesting to formulate it axiomatically.

We have restricted ourselves to working at large central charge in order to both suppress the fluctuation in the spectrum and to avoid off-shell contributions to OPE statistics. Nevertheless, the argument (along with all the expressions presented in this work) is valid at finite $c$. One can therefore also interpret the results as saying that including only hyperbolic manifolds in 3D gravity is a self-consistent model, at least in the sense that it passes the check of crossing symmetry.

Along a different line, the idea of using crossing symmetry to define an ensemble has been employed in the tensor model of \cite{BdBJNS23}. By construction, the model is manifestly crossing symmetric. However, it is still a conjecture that the model produces each 3D manifold exactly once \cite{JRW24}. The statement that crossing symmetry holds if each 3D manifold is counted with weight one, which we demonstrate in this work, is therefore a piece of supporting evidence for the conjecture.

We have focused on closed CFTs. The connection between 3D gravity and CFT ensembles generalizes to BCFTs \cite{WWW25,HJL25}, where the ensemble gains new random variables and the bulk gains new dynamical objects. At finite $c$, the bulk theory is an open-closed extension of Virasoro TQFT \cite{HJL25,JRW25}. A BCFT needs to satisfy six independent bootstrap conditions (crossing symmetries) \cite{Cardy89,CL91}, so it is natural to ask whether the main results of this work generalize to the BCFT ensemble. For example, this will require a generalization of the notation of pseudo-hyperbolicity to manifolds with end-of-the-world branes. It would be interesting to establish this explicitly. 

We have seen that the toroidal surgery can turn a hyperbolic manifold into a pseudo-hyperbolic one and \emph{vice versa}. The same is true also for the half-toroidal surgery where one integrates over the weight of a Virasoro Wilson line extending between two points at asymptotic boundaries \cite{CCHM22,JRW24} (the terminology was introduced in \cite{JRW25}). In fact, neither toroidal nor half-toroidal surgeries could turn a manifold that is either hyperbolic or pseudo-hyperbolic into one that is neither. This suggests that we should treat hyperbolic and pseudo-hyperbolic manifolds on equal footing, at least when dealing with certain questions, such as in the tensor model of \cite{BdBJNS23,JRW24} where some of the interaction vertices are hyperbolic while others are pseudo-hyperbolic.

Finally, it is worth emphasizing again that the spectrum we use in this work, taken from \cite{CCHM22}, is not modular invariant. Upgrading the spectrum to the Maloney-Witten-Keller sum turns the problem of crossing symmetry into a much more difficult one. The argument presented in this work does not straightforwardly generalize. However, interesting progress has been made in \cite{BCELP26}, which constructs an ensemble that has both crossing symmetry and modular invariance by restricting to a particular subset of hyperbolic manifolds.

\section*{Acknowledgements}
It is a pleasure to thank Scott Collier, Tom Hartman, Daniel Jafferis, Liza Rozenberg, Cynthia Yan, and Mengyang Zhang for helpful discussions. In particular, we thank Scott Collier and Tom Hartman for discussions that helped set up the problem of ``enemble crossing symmetry'', and Daniel Jafferis, Liza Rozenberg, and Cynthia Yan for discussions about the treatment of the identity operator. DW acknowledges support by NSF grant PHY-2207659 and the Simons Collaboration on Celestial Holography.

\appendix

\bibliographystyle{JHEP}
\bibliography{lib}
\end{document}